\documentclass[preprint,showpacs,preprintnumbers,amsmath,amssymb]{revtex4}

\usepackage{graphicx}
\usepackage{dcolumn}
\usepackage{bm}
\usepackage{amsmath,dsfont}
\usepackage{amssymb,amsthm,amscd, amsbsy, array}
\input amssym.def
\input amssym.tex

\newcommand{\half}{{\scriptstyle{\frac{1}{2}}}}
\def\2{{\half}}

\def\p{{\partial}}

\def\va{{\bm{a}}}

\newcommand{\vp}{{\bm p}}

\def\beq{\begin{equation}}
\def\eeq{\end{equation}}
\def\beqa{\begin{eqnarray}}
\def\eeqa{\end{eqnarray}}
\def\nn{\nonumber}
\def\barray{\left(\begin{array}}
\def\earray{\end{array}\right)}
\def\barraynb{\begin{array}}
\def\earraynb{\end{array}}
\def\ort{{\rm o}}

\def\smallover#1/#2{\hbox{$\textstyle\frac{#1}{#2}$}} %

\def\vx{{\bm{x}}}

\def\vX{{\bm{X}}}

\def\vnabla{{\bm\nabla}}

\def\half{\frac{1}{2}}
\def\ben{\begin{equation}}
\def\een{\end{equation}}
\def\bea{\begin{eqnarray}}
\def\eea{\end{eqnarray}} 
\def \bx{{\bf x}}
\def \nn{\nonumber}

\begin{document}

\preprint{arXiv: 1112.4793v5 [hep-th]}

\title{Kohn's theorem and Newton-Hooke symmetry
for Hill's equations}

\author{P. M. Zhang$^{1}$\footnote{email:zhpm@impcas.ac.cn},
G. W. Gibbons$^{2}$\footnote{email:G.W.Gibbons@damtp.cam.ac.uk},
P. A. Horvathy$^{1,3}$\footnote{email:horvathy@lmpt.univ-tours.fr}}

\affiliation{$^1$Institute of Modern Physics, Chinese Academy of Sciences
\\
Lanzhou, China \\
$^2$Department of Applied Mathematics and Theoretical  Physics,
Cambridge University, \\ Cambridge, UK\\
$^3$Laboratoire de Math\'ematiques et de Physique
Th\'eorique,
Universit\'e de Tours
\\
Tours, France}

\date{\today}

\begin{abstract}
Hill's equations, which first arose in the study of the
Earth-Moon-Sun system, admit the two-parameter
centrally extended Newton-Hooke symmetry without rotations.
This symmetry allows us to extend Kohn's theorem about the center-of-mass decomposition. Particular light is shed on the problem using Duval's ``Bargmann'' framework.
The separation of the center-of-mass motion into that of a guiding center and relative motion
 is derived by a generalized  chiral decomposition.
\end{abstract}

\pacs{
11.30.-j,   
02.40.Yy,   
02.20.Sv,   
96.12.De,   
\\[8pt]
Phys. Rev. {\bf D85} (2012) 045031
%
%
\\[8pt]
}

\maketitle

\newpage
\tableofcontents
\newpage

\section{Introduction }
The relationship between the ability  to split off
the centre of mass motion, the idea of a ``guiding centre''
and its connection with some form of generalised
Galilei, or Newton-Hooke type  `` kinematic symmetry''
\cite{BacryLeblond,BacryNuyts}  has
been the subject of a number
of recent papers \cite{GiPo,ZH-Kohn,ZH-Kohn-II}.
In the presence of magnetic fields
this is the subject of Kohn's theorem \cite{Kohn} and its variants.
The use of the guiding centre approximation in plasma physics
is well known.  Less well explored is the application of these
ideas to gravitational physics. It is true that
the idea of a guiding centre is well established in galactic
dynamics \cite{Binney}, but its connection with
kinematic symmetries does not appear  to have been explored before.
The purpose of the present paper is to fill that gap in the literature.

The oldest example of what we have in mind are  Hill's equations
for the Earth-Moon-Sun system \cite{HillAJM,Gutzwiller}.
However with the development of our understanding
of the structure of the galaxy, it was realized  that
similar equations hold for the motion of stars
around the Milky Way \cite{Bok,Mineur,Chandra,Binney}.
Understanding many electron atoms in the  Old Quantum  Theory
leads to the same equations and its failure to deal with the Helium
was the notorious stumbling block which led to the development of modern
Quantum Mechanics. In more recent times there has been a revival of
interest in semi-classical models of many electron atoms \cite{Tanner}
and to muonic atoms \cite{Stuchi}.

The plan of the paper is as follows. In section \ref{Hillseqs} we introduce
and derive Hill's equations. In section \ref{SymCOM} we analyze their
 symmetry group and its relation to the centre of mass motion.
and in particular (in the planar case) show that it is
five-dimensional. In section \ref{algebra} we obtain
the Lie algebra using its vector field generators
acting the Newton-Cartan spacetime. In section \ref{momentmap} we pass
to a Hamiltonian treatment and show that the  Poisson algebra
of moment maps is an extension by two central elements.
In section \ref{Eisenhart} we provide the Eisenhart-Duval \cite{Eisen,Duval,DHP2,DHH,DHHP} lift of the
system to a $3+1$ dimensional metric with Lorentz
signature which is not conformally flat,
as we show explicitly. In section
\ref{Landau} we give an alternative interpretation of the Hill system in terms
of a Landau problem in an anisotropic oscillator
and in section \ref{chiral} we use this representation to give a ``Chiral
Decomposition'' using the methods of \cite{ZH-chiral,AGKP}.
Section \ref{variants} describes some possible variants  and extensions
of our results and the last section is a short conclusion.

\section{Hill's equations}\label{Hillseqs}

As a model for the Earth-Moon-Sun system \cite{HillAJM,Gutzwiller}, or for a cluster of stars
moving around the galaxy  in an approximately circular orbit
 \cite{Bok,Mineur,Chandra,Binney}, one has the following equations
\beq\begin{array}{lll}
m_a \bigl(\ddot{x}_a -2\omega \dot{y}_a-3\omega^2 x_a\Bigr)
&=&
\displaystyle\sum_{b\ne a} \frac{Gm_am_b(x_b-x_a)}{|\vx_a-\vx_b|^3}\,,
\\[8pt]
m_a \bigl(\ddot{y}_a + 2 \omega \dot{x}_a \bigr ) &=&
\displaystyle\sum_{b\ne a}\frac{Gm_am_b(y_b-y_a)}{|\vx_a-\vx_b|^3}\,,
\\[8pt]
m_a\bigl(\ddot{z}_a + \omega^2 z_a \bigr) &= &
\displaystyle\sum_{b\ne a}\frac{Gm_am_b
(z_b-z_a)}{|\vx_a-\vx_b|^3}\,.
\end{array}
\label{Hilleqs}
\eeq
These equations are valid in a suitable rotating coordinate system.
For the Earth-Moon-Sun system, for example, $\vx_1$ can be the position of the Earth and $\vx_2$ can be that of the Moon. The motion of the Sun is neglected, and the remnant of its influence on the Earth-Moon pair is represented, in the first-order approximation, by the repulsive anisotropic harmonic term \footnote{The repulsive harmonic potential is tidal in nature. Indeed since it
arises from balancing attractive gravitational and repulsive
centrifugal forces the equilibrium solutions are unstable.} in the first equation, where
\ben
\omega^2 = \frac{GM}{R^3}
\label{KepOmega}
\een
is the angular velocity of a circular Keplerian orbit
lying in the $z=0$ plane and having radius $R$.
The linear-in-velocity terms correspond to the Coriolis force
induced in a rotating coordinate system.
The right hand sides represent the gravitational interactions
between the Earth and the Moon.

These equations are obtained as follows \cite{Heggie,Gurfil}.
Let  $r,\theta^\prime,z$ be  cylindrical  coordinates
centered on the Sun or on the galactic center which has a mass
$M$ so large compared with those  of the other moving masses
that it may be assumed to remain at rest in an inertial
coordinate system. Let $x,y,z$ be coordinates with  respect to
a rotating coordinate system whose origin lies
on the a Keplerian orbit with
$r=R,\, \theta^\prime=\omega t,\, z=0$.
The $x$ axis is taken to be radial  so that
$r= R+x\,,$
and the $y$ axis is taken to be tangential to
the orbit.   The forces acting on each  particle
whose comoving coordinates are  $\vx_a$  consist of  their
mutual gravitational attractions and the attraction
due to the gravitational potential produced by the Galaxy
\ben
U= -\sum_a\frac{GM m_a}{\sqrt{(R+x_a)^2+y_a^2+z_a^2}}.
\een
To quadratic accuracy in $x_a,y_a,z_a$
\ben
U= -\sum_a \frac{GMm_a}{R} \Bigl(1-\frac{x_a}{R}-\half \frac{y_a^2 + z_a^2}{R^2}
+ \frac{x_a^2}{R^2}\Bigr) \,,
\een
which implies that the force is to linear accuracy
\ben
-\nabla U = \sum_a m_a\omega^2 \bigl(-R + 2x_a,-y_a, -z_a \bigr) \,.
\een
Substitution in Newton's equations of motion now gives Hill's equations (\ref{Hilleqs}).

Strictly speaking the equation originally considered
by Hill was a special planar  case for two bodies
in which the position of the Earth
$\vx_1$ was assumed to be at rest $\vx_1=0$ and the position of the Moon
$\vx_2$ thus to satisfy
\beq\begin{array}{llll}
\ddot{x}_2\; - &2\omega \dot{y}_2  -3 \omega ^2 x_2  &=& -\displaystyle\frac{Gm_1 x_2}{(x_2^2 +y_2^2)^{\frac{3}{2}}},
  \nonumber
\\[12pt]
\ddot{y}_2\;  + &2\omega \dot{x}_2 &=& -
\displaystyle\frac{Gm_1 y_2}{(x_2^2+y_2^2)^{\frac{3}{2}}} \,.
\end{array}
\label{UrHill}
\eeq
We omit henceforth the $z$ variables and work in the plane.
We mention however that the more general case where  motion in the
$z$ direction is allowed has important applications, either to the
Earth-Moon-Sun system, see
\cite{Gutzwiller}, or in semi-classical treatments of the helium atom
\cite{Tanner} or muonic atoms \cite{Stuchi}.

\section{Symmetries and center of mass motion}\label{SymCOM}

In addition to the discrete symmetries of parity
 and time reversal,
 \beq
\vx_a(t) \rightarrow -\vx_a(t)
 \quad\hbox{and}\quad
\big(x_a(t),y_a(t)\big) \rightarrow
\big(x_a(-t),-y_a(-t)\big),
\eeq
Hill's  equations admit  a continuous four-parameter  family of  abelian symmetries, since
they are invariant under ``translations and boosts''
\cite{GiPo,ZH-Kohn,ZH-Kohn-II,ZH-chiral},
\beq
\vx_a\to\vx_a+\va(t)\,.
\label{tdtrans}
\eeq
Inserting into the Hill equations and putting (with some abuse of notations)
$\va=(x,y)$, allows us to infer that to be a symmetry requires
\beq
\begin{array}{llll}
\ddot{x}&-\,2\omega\dot{y}-3\omega^2x&=&0
\\[2pt]
\ddot{y}&+\,2\omega\dot{x}&=&0
\end{array}
\label{HillSymmeqs} \,
\eeq

\begin{figure}
\begin{center}
\includegraphics[scale=1]{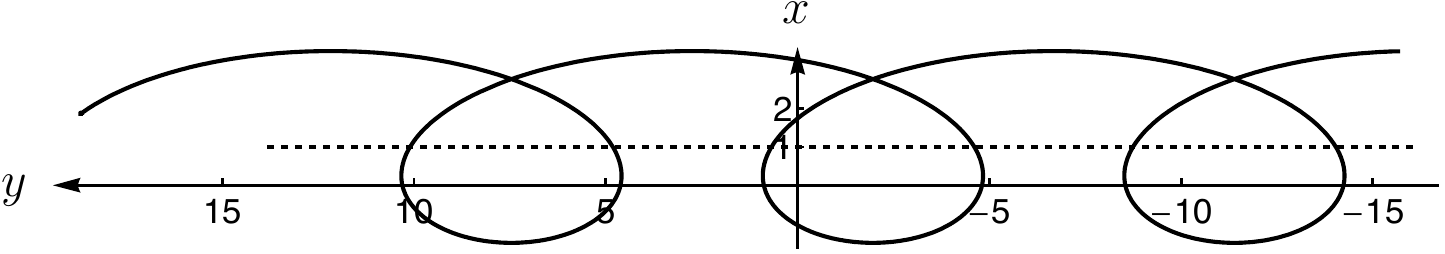}\quad
\vspace{-6mm}
\end{center}
\caption{\it Trajectory of the center of mass in the Hill problem. The straight horizontal line in the middle indicates the trajectory of the guiding center about which the center of mass performs ``flattened elliptic motion''.
}
\label{Hillplot}
\end{figure}

The simplest way to solve these equations is to derive
the upper equation w.r.t. time and then use the lower equation to eliminate $\ddot{y}$ to yield
 an oscillator equation for $\dot{x}$,
${d^2\dot{x}}/{dt^2}=-\omega^2\dot{x}$. Thus
$$
x(t)=\frac{A}{\omega}\sin\omega t-\frac{B}{\omega}\cos\omega t
+x_0.
$$
Putting $x(t)$ into the second equation and integrating provides us with
$y(t)$; testing the pair $x(t),y(t)$ on our original system
fixes the integration constants to yield
\beq\begin{array}{llll}
x(t)&=&\displaystyle\frac{A}{\omega}\sin\omega t -
\displaystyle\frac{B}{\omega}\cos\omega t
&+\;x_0\,
\\[8pt]
y(t)&=&2\displaystyle\frac{A}{\omega}\cos\omega t+
2\displaystyle\frac{B}{\omega}\sin\omega t
&-\;\displaystyle\frac{3}{2}\omega tx_0+y_0.
\end{array}\,.
\label{xysymm}
\eeq
Then
\ben
\frac{\omega^2(x-x_0)^2}{A^2+B^2}+\frac{\omega^2(y-y_0 + \frac{3}{2}\omega x_0 t
  )^2}{4 (A^2+B^2 )} =1
\een
shows that the trajectories are ellipses centered at  $(x_0,y_0-\frac{3}{2}\omega x_0 t)$
with major axes lying along the $y$ direction.
The  ratio of the
semi-major to the semi-minor axis is $2::1$, and
the centers drift along the $y$ direction with constant speed
$\frac{3}{2}\omega x_0$
in the direction of its major axis, see Fig. \ref{Hillplot}.

At this point, it is legitimate to wonder what is the interest of studying
the properties of (\ref{HillSymmeqs}), which only describe
some property of the system but not the physical system itself.
A justification comes from observing that, as a consequence of the linearity of the
l.h.s. of (\ref{Hilleqs}) and  because the gravitational forces on r.h.s. of these equations satisfy Newton's third law, the \emph{center of mass},
\beq
\vX=
\barray{c}x\\ y\earray=
\frac{\sum_{a}m_a\vx_a}{\sum_a m_a},
\label{COM}
\eeq
(with another abuse of notation) satisfies exactly the \emph{same}
equations (\ref{HillSymmeqs}). The latter describes therefore more than a ``property''.

We emphasize that our statement relies on the equality of the inertial and passive gravitational mass, $m_a$,
for objects with significant
self-gravitation. In other words, both
the center-of-mass decomposition
and  Galilean symmetry depend on the so-called \emph{Strong Equivalence Principle} \cite{DHH,ZH-Kohn}.
It has been verified  experimentally by lunar laser ranging to very
high accuracy
using the Nordtvedt effect for the Sun-Earth-Moon system. If
\ben
\frac{m_{\rm passive}}{m_{\rm inertial}} =
1- \eta_N \frac{E_G}{c^2\, m_{\rm inertial} }
\een
 where $ E_G$ is the Gravitational self-energy, then
\ben
|\eta_N| \le (4.4 \pm 4.5) \times 10 ^{-4} \,.
\een
Since $ \frac{E_G}{c^2\, m_{\rm inertial} } =    4.6 \times 10^{-10}$
for the Earth  and $ 2.1 \times 10^{-10}$ for the Moon,
the strong equivalence principle is satisfied to better than one part in $10^{13}$ \cite{Will} .

Below we focus our attention at the symmetry alias center-of mass equation
(\ref{HillSymmeqs}).

The general solution (\ref{xysymm}) is composed of
two particular cases.

$\bullet$ Let us choose first
$x_0=y_0=0$; then the trajectory is
an  ellipse centered at the origin,
and oriented along the $y$ direction,
\beq
\vX_+(t)=\barray{c}
X_+^1(t)\\ X_+^2(t)\earray=
\barray{c}
\displaystyle\frac{A}{\omega}\sin\omega t -
\displaystyle\frac{B}{\omega}\cos\omega t
\\[8pt]
2\displaystyle\frac{A}{\omega}\cos\omega t+
2\displaystyle\frac{B}{\omega}\sin\omega t
\earray\,.
\label{flatell}
\eeq

Putting $A=B=0$ provides us instead with
\beq
\vX_-(t)=\barray{c}
X_-^1(t)\\ X_-^2(t)\earray=
\barray{c}
x_0\\ -\frac{3}{2}\omega tx_0+y_0
\earray.
\label{Hallsol}
\eeq
The particular form of this solution comes from a delicate balance between  the harmonic and the inertial forces which precisely cancel,
\beq
3\omega^2\delta^{1i}X_-^1+2\omega\varepsilon^{ij}\dot{X}_-^j=0,
\label{Halllaw}
\eeq
so that the particle drifts perpendicularly to the harmonic field with constant velocity.
Anticipating what comes below in Sect. \ref{Landau}, we call it a \emph{Hall motion}.

As it will be explained in Sect. \ref{chiral},
the first of these particular solutions, namely
$\vX_-$, describes the \emph{guiding center},
and the second, $\vX_+$,
describes the relative motion around it.

\section{Vector fields and algebra}\label{algebra}

The planar symmetry group is generated by the spacetime vector fields
\beq\begin{array}{lll}
K_+^1&=&\displaystyle\frac{1}{\omega}
\big(\sin\omega t\p_x+2\cos\omega t\p_y\big)
\\[6pt]
K_+^2&=&\displaystyle\frac{1}{\omega}
\big(-\cos\omega t\p_x+2\sin\omega t\p_y\big)
\\[6pt]
K_-^1 &=&\p_x-\displaystyle\frac{3}{2} \omega t \p_y
\\
K_-^2 &=& \p_y
\\[6pt]
H&=& \p _t \,.
\end{array}
\label{svf}
\eeq
Here the vector fields $K_\pm^i,\,i=1,2$ generate the infinitesimal time-dependent symmetries  (\ref{tdtrans}),
and $H$ represents infinitesimal time translations.
The non-trivial brackets are
\beqa\begin{array}{lll}
\bigl[H, K_+^1\bigr] &=& -\omega K_+^2\,,
\\
\bigl[H,K_+^2\bigr] &=& +\omega K_+^1\,,
\\
\big [H,K_-^1\bigr] &=&-\displaystyle\frac{3}{2}\omega K_-^2.
\end{array}
\label{vfLie}
\eeqa

Normalizing the total mass to unity,
$\sum_am_a=1$, the Lagrangian for the center of mass is
\ben
L= \half\bigl(\dot{x}^2+\dot{y}^2\bigr)
-\omega (y\dot{x}-x\dot{y})
+\frac{3}{2}\omega^2x^2.
\label{COMLag}
\een
The mechanical momenta
 $p_x=\dot{x}$ and $p_y=\dot{y}$ do not Poisson-commute,
$
\{p_x, p_y\} = 2 \omega \,.
$
The Hamiltonian is
\ben
H= \half\bigl(p_x^2 + p_y^2
\bigr)-\frac{3}{2}\omega^2x^2
 \,.
\een
It is Liouville integrable since $H$,
 and the dual momentum $\bar p_y=\dot{y}+ 2\omega x$ mutually commute
\cite{Gurfil}.

\section{Moment maps for Hill's equations}\label{momentmap}

Following \cite{GiPo}, we find the conserved quantities
\beq\begin{array}{lll}
\kappa_+^1 &=&\displaystyle\frac{1}{\omega}\big(
\;\;\; p_x\sin\omega t+2p_y\cos\omega t+3\omega x\cos\omega t\big),
 \\[6pt]
\kappa_+^2 &=&\displaystyle\frac{1}{\omega}\big(
 -p_x\cos\omega t+2p_y\sin\omega t+3\omega x\sin\omega t\big),
 \\[6pt]
\kappa_-^1 &=&
 (p_x-\half\omega y)-\displaystyle\frac{3}{2}\omega t(p_y+ 2\omega x),
\\
\kappa_-^2&=&
 p_y+2\omega x\,.
\end{array}
\label{Gcons}
\eeq
Recovering the generating vectorfields in (\ref{svf})
as
\beq
- K=\{\kappa,x\}\p_x+\{\kappa,y\}\p_y
\nn
\eeq
can be viewed as a consistency check.

Note that the Poisson algebra does not coincide with the
Lie algebra (\ref{vfLie}) since $\kappa_+^1$ and $\kappa_+^2$
and $\kappa_-^1$ and $\kappa_-^2$, do not
Poisson commute but  their brackets give rather
\emph{two central extensions},
\beq\begin{array}{llc}
\{\kappa_+^1, \kappa_+^2 \}&=& \;\displaystyle\frac{1}{\omega},
\\[10pt]
\{\kappa_-^1, \kappa_-^2 \}&=& -\displaystyle\half\omega
\,.
\end{array}
\label{2Heis}
\eeq
The other bracket relations (\ref{vfLie}) are unchanged.
Thus the conserved quantities realize two commuting copies of
 Heisenberg  algebras.
Evaluating the moment maps on the solutions (\ref{xysymm})
gives
\beq\begin{array}{llc}
\kappa_+^1&=&\displaystyle\frac{B}{\omega},
\\[6pt]
\kappa_+^2 &=& -\displaystyle\frac{A}{\omega}\,,
\\[6pt]
\kappa_-^1 &=& -\displaystyle\half\omega y_0\,,
\\[6pt]
\kappa_-^2&=& \displaystyle\half\omega x_0 \,,
\end{array}
\label{values}
\eeq
which shows that they are indeed constants of the motion.
Hence $\kappa_-^2=\half\omega x_0 $
 commutes with the Hamiltonian $H$, and is related to the  [conserved] $x$ coordinate
of the center of the revolving ellipse.
In terms of the conserved quantities, the Hamiltonian reads
\ben
H=\frac{\omega^2}{2}\big((\kappa_+^1)^2+(\kappa_+^2)^2\big)-\frac{3}{2}\big(\kappa_-^2\big)^2
= \half(A^2+B^2)-\frac{3}{8}\omega^2x_0^2 \,.
\label{Hamiltonian}
\een
One thus has
\beq\begin{array}{lll}
\{H,\kappa_+^1\} &=& -\omega \kappa_+^2
\\[6pt]
\{H,\kappa_+^2\} &=&+\omega \kappa_+^1
\\[6pt]
\{H,\kappa_-^1\} &=&-\displaystyle\frac{3}{2}\omega \kappa_-^2
\\[6pt]
\{H,\kappa_-^2\} &=& 0\,.
\end{array}
\label{HamKappa}
\eeq
The $\kappa$, although explicitly time-dependent, are however conserved,
$
\displaystyle\frac{d\kappa}{dt}=\displaystyle\frac{\p\kappa}{\p t}
+\big\{\kappa, H\big\} =0\,.
$

\section{Eisenhart-Duval Lift}\label{Eisenhart}

Following the procedure described in
\cite{Duval,DHH,GiPo} we lift the Hamiltonian to that of a massless particle
in 3+1 spacetime dimensions. The calculation is straightforward and we
just give the result. The 4-metric is given by
\beqa
ds^2 &=& dx^2+dy^2 +2dt\bigl(dv +\omega(xdy-ydx)\bigl) + 3 \omega^2x^2dt^2
\label{4metric}
\eeqa
where $v$ is a new, ``vertical'' coordinate \cite{Duval}.

As pointed out in Ref. \cite{DHH}, the $2\omega$-terms
in (\ref{4metric}) admit a two-fold interpretation. In the present context here, they can be viewed as representing inertial forces in our rotating
coordinate system. In Section \ref{Landau} below they will
be interpreted as an external magnetic field.
Its null-geodesics project onto ``ordinary'' spacetime
according to the center-of-mass [alias symmetry] equations
of motion (\ref{HillSymmeqs}).

 (\ref{4metric}) is a
Ricci-flat
 3+1 dimensional Lorentzian
 metric  with covariantly constant
null Killing vector field
\beq
\xi=\p_v,
\eeq
i.e., a ``Bargmann space''  \cite{Duval}.

The ``Bargmann'' framework is particularly convenient for describing the symmetries. Our symmetry  transformations
 lift indeed to
the ``Bargmann'' metric (\ref{4metric}) as \emph{isometries}. Let us assume that (\ref{tdtrans})
satisfies the symmetry condition (\ref{HillSymmeqs}).
Completing (\ref{xysymm}) with
\beq
v\to v-\frac{1}{2}\big(
\va\cdot\dot{\va}+2\dot{\va}\cdot\vx+2\omega\,\va\times\vx
\big),
\label{vlift}
\eeq
a tedious calculation shows that
 the ``Bargmann'' metric (\ref{4metric}) is left invariant.

Working infinitesimally, the vector fields (\ref{svf}) lift as
\beq\begin{array}{lllll}
\widetilde{K}_+^1&=&\displaystyle\frac{1}{\omega}
\big(\sin\omega t\,\p_x+2\cos\omega t\,\p_y\big)
&+
&\big(x\cos\omega t +y\sin\omega t\big)\p_v,
\\[8pt]
\widetilde{K}_+^2&=&\displaystyle\frac{1}{\omega}
\big(-\cos\omega t\,\p_x+2\sin\omega t\,\p_y\big)
&+
&\big(x\sin\omega t-y\cos\omega t\big)\p_v,
\\[8pt]
\widetilde{K}_-^1&=&\p_x-\displaystyle\frac{3}{2}\omega t\,\p_y
&+
&\big(-\displaystyle\frac{3}{2}\omega^2xt+\displaystyle\frac{1}{2}\omega y\big)\p_v,
\\[8pt]
\widetilde{K}_-^2&=&\p_y&+&\omega\,x\,\p_v,
\end{array}
\label{liftedvf}
\eeq
whose Lie brackets are found to be
\beq\begin{array}{lll}
\big\{\widetilde{K}_+^1,\widetilde{K}_+^2\big\}&=&
-\displaystyle\frac{1}{\omega}\,\xi,
\\[10pt]
\big\{\widetilde{K}_-^1,\widetilde{K}_-^2\big\}&=&
\displaystyle\frac{1}{2}\omega\,\xi,
\end{array}
\label{liftcommrel}
\eeq
which are,
 \emph{up to  sign} those, (\ref{2Heis}), satisfied by the associated conserved quantities
\footnote{Duval's starting point
has been to  search for a geometric framework (he calls  ``Bargmann space'')
for the ``Bargmann'' [i.e., the one-parameter central extension of the Galilei] group \cite{Duval}.}.

The lifted symmetries realize hence not
original Lie algebra structure, (\ref{vfLie}), but
rather their central extension with $\xi$, the generator of vertical translations, as central element.

\section{Bargmann spaces with Newton-Hooke symmetry}\label{Weyl}

The origin of  Newton-Hooke symmetry has been understood
a long time ago \cite{Duval}~: the Bargmann space of an isotropic harmonic oscillator with time-dependent spring constant $k(t)$ is
\beq
dx^2+dy^2+2dtdv-k(t)(x^2+y^2)dt^2,
\label{isoOsci}
\eeq
and the massless dynamics ``upstairs''
projects on the oscillator dynamics ``downstairs''.
Newton-Hooke symmetry is represented by the isometries
of this metric, and is indeed a subgroup of
 $\xi=\p_v$-preserving conformal transformations ---
 the latter forming
 the (centrally extended) Schr\"odinger group.

The metric (\ref{isoOsci}) is, furthermore,
Bargmann-conformally flat i.e. can be mapped conformally
onto $4$d Minkowski space by a $\xi$-preserving transformation.

Bargmann-conformally related metrics
share the same
symmetries; a conformally flat Bargmann metric admits
therefore the same [namely Schr\"odinger] symmetry as a free particle.


Now we describe, following Refs. \cite{DHTH,DuvalClaus,DHP2,DHHP}, these
Schr\"odinger-conformally flat spaces.
In $D=d+2>3$ dimensions, conformal
flatness is guaranteed by the
vanishing of the conformal Weyl tensor
\beq
C^{\mu\nu}_{\ \ \rho\sigma}
=
R^{\mu\nu}_{\ \ \rho\sigma}
-
\frac{4}{D-2}\,\delta^{[\mu}_{\ [\rho}\,R^{\nu]}_{\ \sigma]}
+
\frac{2}{(D-1)(D-2)}\,
\delta^{[\mu}_{\ [\rho}\,\delta^{\nu]}_{\ \sigma]}\,R.
\label{Weyltensor}
\eeq
Now $R_{\mu\nu\rho\sigma}\xi^\mu\equiv 0$ for a Bargmann space,
implying some extra conditions
on the curvature. Inserting the identity
$\xi_\mu R^{\mu\nu}_{\ \ \rho\sigma}=0$
into
$C^{\mu\nu}_{\ \ \rho\sigma}=0$,
using the identity
$\xi_\mu R^\mu_\nu\equiv 0$
($R^\nu_\sigma\equiv R^{\mu\nu}_{\ \ \mu\sigma}$),
we find
$$
0=
-\left[
\xi_\rho R^\nu_\sigma-\xi_\sigma R^\nu_\rho\right]
+
\frac{R}{D-1}\left[
\xi_\rho\delta^\nu_\sigma-\xi_\sigma\delta^\nu_\rho
\right].
$$
Contracting again with $\xi^\sigma$ and using that $\xi$ is null,
we end up with
$R\,\xi_\rho\xi^\nu=0$. Hence the scalar curvature vanishes, $R=0$.
Then the previous equation yields
$\xi^{ }_{[\rho}R^\nu_{\sigma]}=0$ and thus
$R_\sigma^\nu=\xi_\sigma\eta^\nu$
for some vector field $\eta$. Using the
symmetry of the Ricci tensor, $R_{[\mu\nu]}=0$,
we find that $\eta=\varrho\,\xi$
for some function $\varrho$. We finally get the  consistency relation
\beq
R_{\mu\nu}=\varrho\,\xi_\mu\xi_\nu.
\label{NCfieldeq}
\eeq
The Bianchi identities ($\nabla_\mu R^\mu_\nu=0$
since $R=0$) yield $\xi^\mu\partial_\mu\varrho=0$, i.e. $\varrho$ is
a function on spacetime.
The conformal Schr\"odinger-Weyl tensor is hence of the
form
\beq
C^{\mu\nu}_{\ \ \rho\sigma}=
R^{\mu\nu}_{\ \ \rho\sigma}
-
\frac{4}{D-2}\,
\varrho\,\delta^{[\mu}_{\ [\rho}\,\xi^{\nu]}\xi^{ }_{\sigma]}.
\label{SchWeyl}
\eeq
It is noteworthy that Eq.~(\ref{NCfieldeq}) is
the Newton-Cartan field equation with
$\varrho/(4\pi G)$ as matter density.
 Eq.~(\ref{NCfieldeq}) also implies that the transverse Ricci tensor of a  Schr\"odinger-conformal flat Bargmann metric necessarily vanishes, $R_{ij}=0$ for each~$t$.

Further results are only worked out for total Bargmann dimension $D=4$.
Since the transverse space is  $d=2$-dimensional,
$R_{ij}=0$ implies that
the latter is (locally) flat and we
can choose $g_{ij}=g_{ij}(t)$.
Then a change of coordinates
$(\vx,t,v)\to(G(t)\,\vx,t,v)$ where
$G=(G_{ij})$ is the square-root
matrix $\delta_{ab}G_{i}^aG_{j}^b=g_{ij}$,
casts our Bargmann metric into the form
\beq
dx^2+dy^2+2dt\big[dv+{{\cal A}}\cdot d{\vx}\big]-2Udt^2.
\label{genBmetric}
\eeq

Now we turn to determining all such conformally flat $4$-metrics.
The non-zero components of the Weyl tensor
of (\ref{genBmetric}) are
\begin{eqnarray*}
C_{xyxt}&=&-C_{ytts}=-\frac{1}{4}\partial_x{B},
\\[6pt]
C_{xyyt}&=&+C_{xtts}=-\frac{1}{4}\partial_y{B},
\\[6pt]
C_{xtxt}&=&-\2\big[
\partial_t(\partial_y{\cal A}_y-\partial_x{\cal A}_x)-
{\cal A}_x\partial_y{B},
\big]
+\2\big[\partial_x^2-\partial_y^2\big]U,
\\[6pt]
C_{ytyt}&=&\2\big[
\partial_t(\partial_y{\cal A}_y-\partial_x{\cal A}_x)
-{\cal A}_y\partial_x{B}\big]
-\2\big[\partial_x^2-\partial_y^2\big]U,\\[6pt]
C_{xtyt}&=&\2\big[
\partial_t(\partial_x{\cal A}_y+\partial_y{\cal A}_x)
+2\partial_x\partial_yU\big]
-\smallover1/4({\cal A}_x\partial_x-{\cal A}_y\partial_y){B}.
\end{eqnarray*}
Then Schr\"odinger-conformal flatness requires
\beq
\left\{\begin{array}{lll}
{\cal A}_i&=&\2\epsilon_{ij}{B}(t)x^j+a_i,
\qquad
\vnabla\times\va=0,
\qquad\partial_t\va=0,
\\[6pt]
U(t,\vx)&=&\2 C(t)r^2+{\bm F}(t)\cdot\vx+K(t).
\end{array}\right.
\label{flatmetric}
\eeq
Note, {\sl en passant}, that Eq. (\ref{NCfieldeq}) automatically holds
in this case, because
\beq
R_{xt}=2C_{xtts}=0\,,
\qquad
R_{yt}=2C_{ytts}=0.
\eeq
The only non-vanishing component of the Ricci tensor is
\beq
R_{tt}=
-\partial_t(\vnabla\cdot{\cal A})-\2{B}^2-\Delta U =-\frac {1}{2}B(t)^2 -2C(t).
\eeq

The  metric (\ref{flatmetric}) describes a
uniform magnetic field ${B}(t)$,
an [attractive or repulsive, $C(t)=\pm\omega^2(t)$]
\emph{isotropic} oscillator and a uniform force field ${\bm F}(t)$ in the plane.
All fields may depend arbitrarily on time. It also includes a curl free
vector potential~$\va(\vx)$ that can be gauged away if the
transverse space is simply connected: $a_i=\partial_if$ and the coordinate
transformation $(t,\vx,v)\to(t,\vx,v+f)$
results in the `gauge'
transformation
${\cal{A}}_i\to{\cal{A}}_i-\partial_if=-\2{B}\,\epsilon_{ij}x^j$.
If, however, space is not simply connected, we can
also include an external Aharonov-Bohm-type vector potential,
explaining the $\ort(2,1)$ conformal symmetry of a magnetic vortex \cite{JackiwMV}.

Our "one-sided"
anisotropic oscillator here does \emph{not} qualify therefore: its Weyl tensor does not vanish due to the anisotropy:
\beq
\big[\p_x^2-\p_y^2\big]U\neq0
\quad\hbox{for}\quad
U(x,t)=-\frac{3}{2}\omega^2x^2.
\label{nonflat}
\eeq
The 4-metric (\ref{4metric}) \emph{can not be mapped conformally to empty Minkowski space.}

How can it have the ``Newton-Hook-type'' symmetry, then ?
There is no contradiction, though.
Let us stress that we did \emph{not} find here full Newton-Hooke
symmetry, only its  time dependent translational part~: \emph{rotational symmetry} is plainly broken
for the metric (\ref{4metric}). The latter does \emph{not}
come therefore by ``importing'' from the free case.

\section{Relation to the Landau problem}\label{Landau}

Now we point out that the center-of-mass Hill system (\ref{HillSymmeqs}) can also be be viewed as
a charged anisotropic harmonic oscillator in a uniform magnetic field described by the planar Hamiltonian system
\beqa
\left\{x^i,x^j\right\}=0,
\;
\left\{x^i,p^j\right\}=\delta^{ij},
\;
\left\{p^i,p^j\right\} =eB\,\varepsilon^{ij},
\qquad
H=\frac{\vp^2}{2}+\frac{k_1}{2}(x^1)^2+\frac{k_2}{2}(x^2)^2,
\quad
\label{BOscPH}
\eeqa
where we still scaled the total mass to unity.
Comparing the equations of motion
$
\dot{\xi}=\left\{\xi,H\right\}
$ implying
$
\ddot{x}^i-{eB}{}\epsilon^{ij}\dot{x}^j
+{k_i}{}x^i=0
$
[no sum on $i$ in the last term],
with (\ref{HillSymmeqs}) shows that the Hill system can indeed be viewed as  a repulsive  anisotropic oscillator in a uniform magnetic background with
\beq
k_1\equiv k=-3\omega^2,
\qquad
k_2=0,
\qquad
eB=2\omega.
\label{matching}
\eeq
The identity of the two systems relies on the equivalence of the inertial Coriolis  force in a rotating frame with the Lorentz force due to an external magnetic field  \cite{DHP2,DHH}.
Note also that the condition (\ref{Halllaw}) is then in fact
the Hall law,
\beq
\dot{X}_-^i=\varepsilon^{ij}\frac{E^j}{B}
\label{Halllawbis}
\eeq
with the identifications $eE^i=3\omega^2\delta^{1i}X_-^1$ and $eB=2\omega$.

Let us also stress that the possibility of decomposing the magnetic field  plus oscillator system into
 center-of-mass plus relative motion depends on Galilean \cite{SSD,ZH-Kohn}
(more precisely, on Newton-Hooke \cite{GiPo,ZH-Kohn-II,NHlit}) symmetry,
 which  requires in turn the Kohn condition charge/mass = constant to hold. Furthermore, as  ``charge'' equals ``mass''
here, Kohn's condition is automatically satisfied, providing us with the required Galilean symmetry.

In Section \ref{chiral}, the system will be further analyzed by
decomposing  (\ref{HillSymmeqs}) into chiral components
along the lines of \cite{AGKP} as adapted to the Landau problem \cite{Banerjee,ZH-chiral}.

\section{Chiral decomposition of the Hill system}\label{chiral}

The problem can further be analyzed by decomposing
 our magnetic field + anisotropic oscillator [$k_1=k,\, k_2=0$]
 system into chiral components, generalizing the trick of Ref. \cite{AGKP,ZH-chiral}.
Define indeed the two planar vectors
$\vX_\pm=(X_\pm^i)$ as
 \footnote{The upper indices are vector indices, not powers.}
\beq
p^1=\alpha_{+}X_{+}^2+\alpha_{-}X_{-}^2,\qquad
p^2=-\beta_{+}X_{+}^1-\beta _{-}X_{-}^1,
\qquad
\vX=\vX_{+}+\vX_{-},
\eeq
where $\alpha_\pm$ and $\beta_\pm$ are suitable coefficients to be found.
The symplectic form
\beq
\Omega=dp^i\wedge dx^i+\frac{eB}2\varepsilon^{ij}dx^i\wedge dx^j,
\eeq
 whose associated Poisson bracket is (\ref{BOscPH}),
is  written as
\begin{eqnarray*}
\Omega  &=&\left(-\alpha_{+}-\beta_{+}+eB\right) dX_{+}^1\wedge dX_{+}^2
+\left(-\alpha_{-}-\beta_{-}+eB\right)
dX_{-}^1\wedge dX_{-}^2
\\[8pt]
&&
+\;\Big\{\left(-\alpha _{-}-\beta_{+}+eB\right)
dX_{+}^1\wedge dX_{-}^2
+\left( \alpha _{+}+\beta _{-}-eB\right)
dX_{+}^2\wedge dX_{-}^1\Big\}\,.
\end{eqnarray*}
The symplectic form splits into two uncoupled ones when
\begin{equation}
\alpha_{-}+\beta_{+}=eB,
\qquad
\alpha _{+}+\beta _{-}=eB.
\label{OmegaSplitcond}
\end{equation}

The Hamiltonian becomes in turn
\begin{eqnarray*}
H &=&\frac 1{2}\left( \alpha_{+}^2X_{+}^2X_{+}^2+\beta
_{+}^2X_{+}^1X_{+}^1+\alpha_{-}^2X_{-}^2X_{-}^2+\beta
_{-}^2X_{-}^1X_{-}^1\right)+\frac{k}{2}\left(
X_{+}^1X_{+}^1+X_{-}^1X_{-}^1\right)
\\[8pt]
&&+\;\Big\{\left(\alpha_{+}\alpha_{-}\right)X_{+}^2X_{-}^2
+\left(\beta_{+}\beta_{-}+k\right) X_{+}^1X_{-}^1\Big\},
\end{eqnarray*}
which splits into $H=H_++H_-$ when
\begin{equation}
\alpha_{+}\alpha_{-}=0,
\qquad
\beta_{+}\beta _{-}+k=0.
\label{HamSplitcond}
\end{equation}
Since our formulae are symmetric in $\alpha_+$ and $\alpha_-$, we can
choose $\alpha_-=0$ to find
\beq
\alpha_-=0,\qquad
\alpha_{+}=eB+\frac{k}{eB},
\qquad
\beta_{+}=eB,\;\;\;\;\beta_{-}=-\frac{k}{eB}.
\end{equation}
With such a choice we will have decomposed our system as
\beqa
\Omega&=&
\underbrace{-\big(eB+\frac{k}{eB}\big)
dX_{+}^1\wedge dX_{+}^2}_{\Omega_+}
\;+\;\underbrace{
\big(eB+\frac{k}{eB}\big)
dX_{-}^1\wedge dX_{-}^2}_{\Omega_-}
\label{osciOsplit}
\\[8pt]
H&=&
\underbrace{
\frac{1}{2}\Big[
eB\big(eB+\frac{k}{eB}\big)X_+^1X_+^1
+(eB+\frac{k}{eB})^2X_+^2X_+^2\Big]}_{H_+}\;
+\underbrace{\frac{k}{2eB}\left(eB+\frac{k}{eB}\right)
X_-^1X_-^1
\Big]}_{H_-}.\qquad
\label{osciHsplit}
\eeqa
Note that $\Omega_+$ and $\Omega_-$ have opposite signs.

Returning to the Hill problem,
inserting  the matching coefficients (\ref{matching})
into (\ref{osciOsplit})-(\ref{osciHsplit}) yields
\beq
p^1=\frac{1}{2}\omega X_{+}^2,
\qquad
p^2=-2\omega X_{+}^1-\displaystyle\frac{3}{2}\omega X_{-}^1,
\qquad
\vX=\vX_{+}+\vX_{-},
\eeq
and hence
\beqa
\Omega&=&\Omega_++\Omega_-=
\Big\{-\frac{1}{2}\omega\, dX_{+}^1\wedge dX_{+}^2\Big\}
+
\Big\{\frac{1}{2}\omega\, dX_{-}^1\wedge dX_{-}^2\Big\}\,,
\\[12pt]
H&=&H_++H_-=
\Big\{\frac{1}{2}\omega^2X_{+}^1X_{+}^1+
\frac{1}{8}\omega^2X_{+}^2X_{+}^2\Big\}
\;-\;
\Big\{\frac{3}{8}\omega^2\,X_{-}^1X_{-}^1\Big\} .
\eeqa
The Poisson brackets associated to this symplectic form
show that both sets of
coordinates $X_+^i$ and $X_-^i$ are non-commuting,
\begin{equation}
\left\{X_{+}^1,X_{+}^2\right\} =\frac{2}{\omega},
\quad
\left\{X_{+}^1,X_{-}^2\right\} =\left\{X_{+}^2,X_{-}^1\right\}
=0,
\quad
\left\{X_{-}^1,X_{-}^2\right\}=-\frac{2}{\omega}\,,
\end{equation}
and provide us with the separated equations of motion
\beqa\begin{array}{lllllll}
\dot{X}_{+}^1&=&\displaystyle\frac{1}{2}\omega X_{+}^2,
&\null\qquad
&\dot{X}_{+}^2&=&-2\omega X_{+}^1,
\\[8pt]
\dot{X}_{-}^1&=&0,
&\null\qquad
&\dot{X}_{-}^2&=&-\displaystyle\frac{3}{2}\omega X_{-}^1,
\end{array}
\label{chiraleq}
\eeqa
whose solution allows us to recover
(\ref{flatell})-(\ref{Hallsol})
 once again. The general solution (\ref{xysymm})
is the sum of the chiral components,
$\vX(t)=\vX_+(t)+\vX_-(t)$.

Here the simple $\vX_-$ dynamics is that of Hall motion with constant
velocity drift (\ref{Hallsol}) of that of the \emph{guiding center}, while
the $\vX_+$ system, whose trajectories are those flattened ellipses in (\ref{flatell}) describes the anisotropic oscillations of
 the center of mass about the guiding center.

Having decomposed the center-of-mass alias time-dependent translation-symmetry equation into chiral components, the Newton-Hook symmetry plainly follows from those of
our chiral solutions.
For the separated equations (\ref{chiraleq}) the inital conditions,
\beqa
\vX_+(0)&=&\barray{c}
-B/\omega\\[6pt] 2A/\omega
\earray=\barray{c}X_+^1(t)\cos\omega t-
\half X_+^2(t)\sin\omega t
\\[6pt]
 2X_+^1(t)\sin\omega t
+X_+^2(t)\cos\omega t\earray
\\[16pt]
\vX_-(0)&=&\barray{c}
x_0\\[6pt] y_0\earray
=
\barray{c}
X_-^1(t)
\\[6pt]
X_-^2(t)+\displaystyle\frac{3}{2}\omega tX_-^1(t),
\earray
\eeqa
are plainly constants of the motion. They are in fact proportional to those in Eqs. (\ref{Gcons}) and (\ref{values}),
\beqa
\vX_+(0)=\barray{c}
-\kappa_+^1\\
-2\kappa_+^2\earray,
\qquad
\vX_-(0)=\barray{c}
2\kappa_-^2
/\omega
\\-2\kappa_-^1
/\omega
\earray\,.
\eeqa

\section{Some Variants}\label{variants}

The ideas of the present paper may be generalized
in various directions and
and even applied to areas beyond the
realm of classical gravity to quantum semi-classical
treatments of quantum systems. In this section we
we briefly outline some examples.

\subsection{Anisotropy}
The application of Hill's equations
to galactic clusters
was first suggested by Bok \cite{Bok} and by Mineur \cite{Mineur}
and developed by Chandrasekhar \cite{Chandra}.
Chandrasekhar did not assume that the gravitational field
of the galaxy was just a  simple monopole.
As a result he obtained equations of a more general form:
\beq\begin{array}{lll}
m_a\bigl(\ddot{x}_a-2\omega \dot{y}_a-3\omega_1^2x_a\Bigr)
&=&\displaystyle\sum_{b\ne a}
\displaystyle\frac{Gm_am_b(x_b-x_a)}{|\bx_a-\bx_b|^3}
\\[10pt]
m_a\bigl(\ddot{y}_a + 2\omega \dot{x}_a \bigr ) &=&
\displaystyle\sum_{b\ne a}
\displaystyle\frac{Gm_am_b(y_b-y_a)}{|\bx_a-\bx_b|^3 }
\\[10pt]
m_a\bigl(\ddot z_a+\omega_3^2 z_a\bigr) &= &
\displaystyle\sum_{b\ne a}
\displaystyle\frac{Gm_am_b
(z_b-z_a)} {|\bx_a-\bx_b|^3 }
\end{array}
\label{Hill2}
\eeq
where the $z$ coordinates were restored.
One still has the abelian symmetry (\ref{tdtrans})
but
(\ref{HillSymmeqs}) becomes
\beq\begin{array}{lll}
\ddot{x}-2\omega \dot{y} - 3\omega_1^2 x &=0&
\\
\ddot{y} + 2\omega\dot{x} &=0&
\\
 \ddot z +\omega_3^2 z  &=0&
\end{array}
\eeq
and  (\ref{xysymm}) is replaced by
\beqa
x&=&\frac{A}{\Omega}\sin\Omega t-\frac{B}{\Omega}\cos\Omega
t+x_0,
\nonumber \\[8pt]
y&=& 2A\frac{\omega}{\Omega^2}
\cos\Omega t + 2B\frac{\omega}{\Omega^2}\sin\Omega t
 -\frac{3\omega_1^2}{2\omega }x_0t + y_0,
 \nonumber \\[8pt]
z&=&C \cos \omega_3 t + D \sin \omega_3 t
\,,
\label{newsols}
\eeqa
where $\Omega=\sqrt{4\omega^2-3\omega_1^2}$ is called the {\it epicyclic
frequency}  and often denoted by $\kappa$.
The ellipses now have ratio of major to minor axis
equal to  $\frac{2\omega}{\Omega} $ and move with speed
$\frac{3 \omega_1^2}{2 \omega} x_0 $.

The symmetry is generated by the vector fields
\beq\begin{array}{lll}
K_1&=& \sin\Omega t\,\p_x +
\displaystyle\frac{\omega}{\Omega}
2\cos \Omega  t\,\p_y
\\[8pt]
K_2&=&2\displaystyle\frac{\omega}{\Omega}\sin \Omega t\,\p_y - \cos\Omega t\,\p_x
\\[8pt]
K_3&=& \p_x-\displaystyle\frac{3\omega_1^2}{2\omega} t \,\p_y
\\[8pt]
K_4 &=& \p_y
\\[8pt]
K_5 &=& \cos\omega_3 t\,\p_z
\\[8pt]
K_6&=& \sin\omega_3 t\,\p_z
\\[8pt]
H&=&\p_t \,
\end{array}
\eeq
whose non-trivial brackets read
\beq\begin{array}{lll}
\bigl[H, K_1\bigr] &=& -\Omega K_2
\\[6pt]
\bigl[H, K_2 \bigr] &=& +\Omega K_1
\\[6pt]
\big[H, K_3 \bigr] &=&-\displaystyle\frac{3\omega_1^2}{2\omega}K_4
\\[6pt]
\big[H, K_5 \bigr] &=& -\omega_3K_6
\\[6pt]
\big[H, K_6\bigr] &=&\omega_3K_5.
\end{array}
\label {Lie2}
\eeq
By rescaling the generators,
the sub-algebra they span may  be seen to be independent
of the parameters $\omega$ and $\omega_1$.

\subsection{Electromagnetic Variant}

We could consider a very heavy, and hence immobile,  nucleus
of charge $Ze$ around which electrons of mass $m$ and charge $-e$
move. The equations of motion would be identical to those
in (\ref{Hilleqs}) but the charges on the r.h.s. replacing the masses and the angular velocity becoming now
$\omega^2 = \displaystyle\frac{Ze^2}{mR^3} \,.$

This idea has been exploited in atomic physics.
The most basic example being semi-classical treatments of the Helium
atom \cite{Tanner}. The idea also extends to muonic atoms
\cite{Stuchi}.

\subsection{Time dependence}

Oh et al. \cite{OhLinA}, and Heggie et al. \cite{Heggie,Heggie2}  mention, in the context of
galactic dynamics,  a time dependent
version of (\ref{Hilleqs}) in which the radius $R$ is allowed to
depend on time. In the planar case they are
\beqa
m_a\bigl(\ddot{x}_a-2\omega\dot{y}_a-
(3\omega^2 - 2\frac{\ddot R}{R})x_a+
2 \omega\frac{\dot R}{R}y_a \bigr)
&=&\sum_{b\ne a}\frac{Gm_am_b(x_b-x_a)}{|\bx_a-\bx_b|^3}
\nonumber  \\
m_a \bigl(\ddot{y}_a + 2\omega \dot{x}_a -\frac{\ddot R} {R} y  \bigr   ) &=& \sum_{b\ne a} \frac{Gm_am_b(y_b-y_a)} {|\bx_a-\bx_b|^3 }
\label{Hilltime} \,.
\eeqa

Generalized Galilei invariance (\ref{tdtrans}) still holds
but (\ref{HillSymmeqs}) becomes
\beqa
\ddot{x}-2\omega\dot{y}-(3\omega^2-2 \frac{\ddot R}{R} )     x
+ 2 \omega \frac{\dot R}{R} y &=0,&
\nonumber \\[6pt]
\ddot{y} + 2 \omega \dot{x}  -\frac{\ddot R}{R} y   &=0\,.&
\label{eqstime}
\eeqa

For given $R(t)$ this has a four parameter family of solutions
but if one works out the Killing vector fields
and takes the bracket with time translations $\p_t$
the algebra  will not, in the generic case, close
on a finite dimensional Lie algebra, even if one adds additional generators. The situation is
 reminiscent to the one considered in Refs. \cite{GomLuk,LiuTian,DHCG}. Details will be presented elsewhere.


\section{Conclusion}

A remarkable aspect of Hill's equations is
that our Eqn. (\ref{HillSymmeqs}),
simultaneously describes time-dependent symmetries
(\ref{tdtrans}) and the motion of the center of mass.
Our solutions (\ref{xysymm})
represent therefore the trajectories both of the symmetry
group acting on space-time, and of the center-of-mass.

As long as we consider the 3-body problem
 it would be physically more important to study the relative-motion equation (\ref{UrHill});
 the center of mass has little interest, for, say, Lunar
 motions. But Hill's equations also arise when describing an electron beam in a synchrotron;
guiding center motion is plainly interesting
 for the latter, as it is in plasma physics,
 or in stellar dynamics \cite{Heggie2}.

As explained in \cite{GiPo,ZH-Kohn,ZH-Kohn-II}, the ability to split off the centre of mass motion
relies on Galilean (in fact Newton-Hooke) symmetry.
In our case here rotations are broken, but our time-dependent symmetries suffice.

Moreover, the motion of the center-of-mass can further be decomposed into that of the guiding center and
relative motion; the generalization of the chiral
decomposition \cite{AGKP,Banerjee,ZH-chiral} is ideally
suited for that. It is
worth mentioning that, as in
\cite{ZH-Kohn-II,ZH-chiral}, our calculations could be extended
to ``exotic'' [i.e. non-commutative] particles.

\begin{acknowledgments}
P.A.H and P.-M. Z. acknowledge for hospitality  the \textit{
Institute of Modern Physics} of the Lanzhou branch of
the Chinese Academy of Sciences, and the \textit{Laboratoire de Math\'ematiques et de Physique Th\'eorique} of Tours University, respectively.
We would like to thank C. Duval, A. Galajinsky, J. Gomis,
M. Plyushchay and C. Pope for their interest and for
discussions and correspondence.
 This work was  partially supported by the National Natural Science Foundation of
China (Grant No. 11035006 and 11175215) and by the Chinese Academy of Sciences visiting
professorship for senior international scientists (Grant No. 2010TIJ06).
\end{acknowledgments}


\end{document}